\documentclass[3p,times,twocolumn]{elsarticle}

\usepackage{ecrc}

\volume{00}

\firstpage{1}

\journalname{Nuclear and Particle Physics Proceedings}

\runauth{S.~Y.~Chen, B.~W.~Zhang and E.~Wang}

\jid{nppp}

\jnltitlelogo{Nuclear and Particle Physics Proceedings}

\usepackage{amssymb}

\usepackage[figuresright]{rotating}

\begin{document}

\begin{frontmatter}



\dochead{}

\title{Medium modification of averaged jet charge in heavy-ion collisions}


\author[label1]{Shi-Yong Chen}
\author[label1]{Ben-Wei Zhang}
\author[label1]{Enke Wang}
\address[label1]{Key Laboratory of Quark \& Lepton Physics (MOE) and Institute of Particle Physics,
 Central China Normal University, Wuhan 430079, China}



\begin{abstract}

Jet charge characterizes the electric charge distribution inside a jet. In this talk we make the first theoretical study of jet charge in high-energy nuclear collisions and calculate numerically the medium alternations of jet charge due to parton energy loss in the quark-gluon plasma. The parton multiple scattering in hot/dense QCD medium is simulated by a modified version of  PYQUEN Monte Carlo model with 3+1D ideal hydrodynamical evolution of the fireball. Our preliminary results show that the averaged jet charge is significant modified in A+A collisions relative to that in p+p. The different features of quark jet charge and gluon jet charge in heavy-ion collisions, and the sensitivity of jet charge modifications to flavour dependence of energy loss are observed, which could then be used  to discriminate quark and gluon jet as well as their energy loss patterns in heavy-ion collisions.
\end{abstract}


\end{frontmatter}

\section{Introduction}
\label{Introduction}




One very important purpose of high energy heavy-ion collisions performed at RHIC
and LHC is to study the creation and  the properties of Quark-Gluon Plasma (QGP),
a new form of matter with deconfined quarks and gluons created during these collisions.  It has been shown that the jet quenching, or the energy loss of a fast parton traversing the nuclei could be used a very sensitive probe to calibrate the density and temperature  of the QCD matter~\cite{Gyulassy:2003mc}. The attenuation of parton energy loss will engender the suppression of leading hadrons~\cite{Gyulassy:2003mc,Dai:2015dxa}  as well as alternations of full jet observables in the high-energy nuclear collisions  such as jet shapes and fragmentation functions~\cite{Vitev:2008rz,Qin:2015srf}, single jet cross section~\cite{Vitev:2009rd}, di-jet transverse momentum asymmetry~\cite{He:2011pd} and tagged jet $p_{T}$ imbalance~\cite{Dai:2012am}, {\it etc}.

With the advance of expereimental measurements and theoretical calculations of jet quenching, it is of importance to explore the jet quenching effect on new full jet observable with distinct features.
A very interesting jet observable is jet charge, which gives the electric charge distribution in a recosntructed jet and was firstly proposed by Field and Feynman almost forty years ago~\cite{Field:1977fa}. This observable has been investigated experimentally in deep inelastic scattering studies to understand
the correlation between the quark model and hadrons. Since then, jet charge observables have
been utilized in tagging the charge of b-quark jets, hadronically
decaying $W$ bosons,  and distinguishing quark jets from gluon jets~\cite{Krohn:2012fg,Waalewijn:2012sv}.
Recently, ATLAS Collaboration published their measurement of jet charge in dijet events in
$\sqrt{s}=8$~TeV p+p collisions~\cite{Aad:2015cua}. In this talk we make a first theoretical investigatation of jet charge in relativistic heavy-ion collisions and see its medium modification in relative to hadronic collisions~\cite{Chen:2016}. We show that parton energy loss of the energetic partons in the QGP will change the electric charge distributions in a full jet and demonstrate the obvious distinction between quark and gluon jet charges, which can be used as a excellent tool to study the different patterns of energy loss for quarks and gluons~\cite{Liu:2006sf, Chen:2008vha}.

The paper is organized as follows. In Section \ref{sec:theory} we discuss jet charge at p+p collisions and then show the theoretical framework of computing jet charge in A+A reactions including jet quenching effect. In Section \ref{sec:results} the preliminary results of medium modification of jet charge are provided with discussions. At last, we give a brief summary in Section \ref{sec:summary}.

\section{Theoretical framework}
\label{sec:theory}

The momentum-weighted electric charge sum of hadrons in the jet,
called the averaged jet charge, is defined as:
\begin{eqnarray}
Q^{i}_{\kappa}=\sum_{ h \ \in jet}z^{\kappa}_{h}Q_{h}
\label{TAB}
\end{eqnarray}
where $i$ indicates the jet flavour, and $z_h$ is the momentum fraction of the jet carried by the fragmentated hadron $h$ with electric charge $Q_{h}$. $\kappa$ is a power parameter  which varies from 0.1 and 1 to adjust the bias of hadrons with different transverse momentum.

\begin{figure}[!htb]
\centerline{
\includegraphics[width = 70mm, height=2.5in,angle=0]{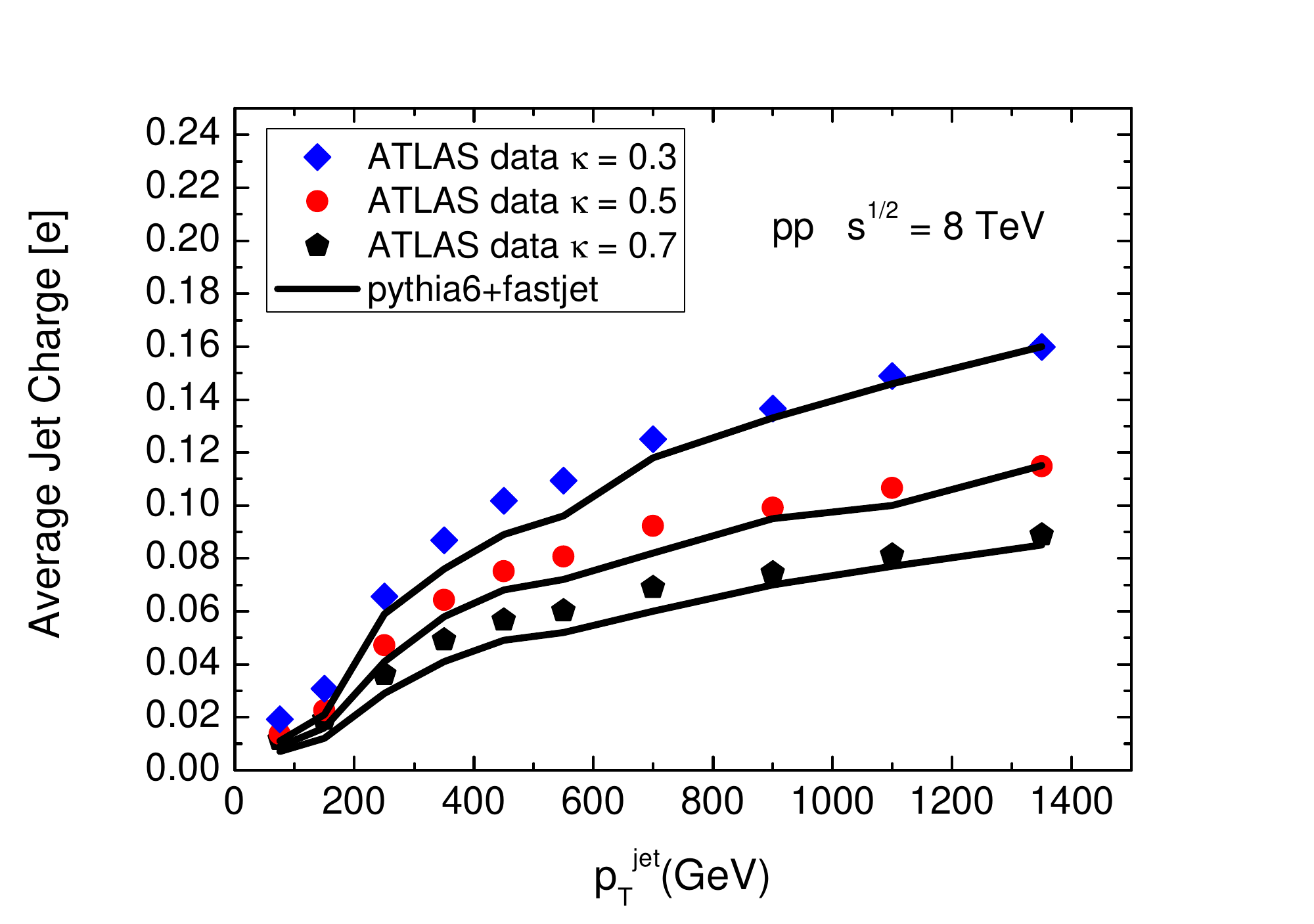}
}
\caption{The computed average jet charges for the leading jet
in dijet events as a function of the jet transverse momentum. The experiment data by ATLAS have also been shown for a comparison.}
\label{fig:ATLAS}
\end{figure}

Firstly we study jet charge in p+p collisions. It has been shown that
the energy and and jet-size dependence of moments of jet-charge distributions can be
calculated in perturbative QCD in~\cite{Krohn:2012fg}.
here, we employ the MC event generater pythia6
with the Perugia 2012 tune~\cite{Skands:2010ak} to
simulate particle production in p+p collisions, and the FastJet package~\cite{Cacciari:2008gp} is used to construct final state jets.
Shown in in Fig.~\ref{fig:ATLAS} are the results for jet charge in p+p collisons at $\sqrt{s}=8$~TeV from {\sc pythia6+fastjet}
compared with ATLAS recent data~\cite{Aad:2015cua}. One can observe that jet charge goes up with increasing jet transverse momentum, and the distributions with larger $\kappa$ are higher~\cite{Chen:2016}.
Decent agreement can be seen between experimental data and the theoretical calculation.  In the following we use the same approach to predict jet charge
at $\sqrt{s}=2.76$~TeV p+p collisions, which may provide a reasonable baseline for calculating jet charge in heavy-ion collisions.

\begin{figure}[!htb]
\centerline{
\includegraphics[width =70mm, height=2.3in,angle=0]{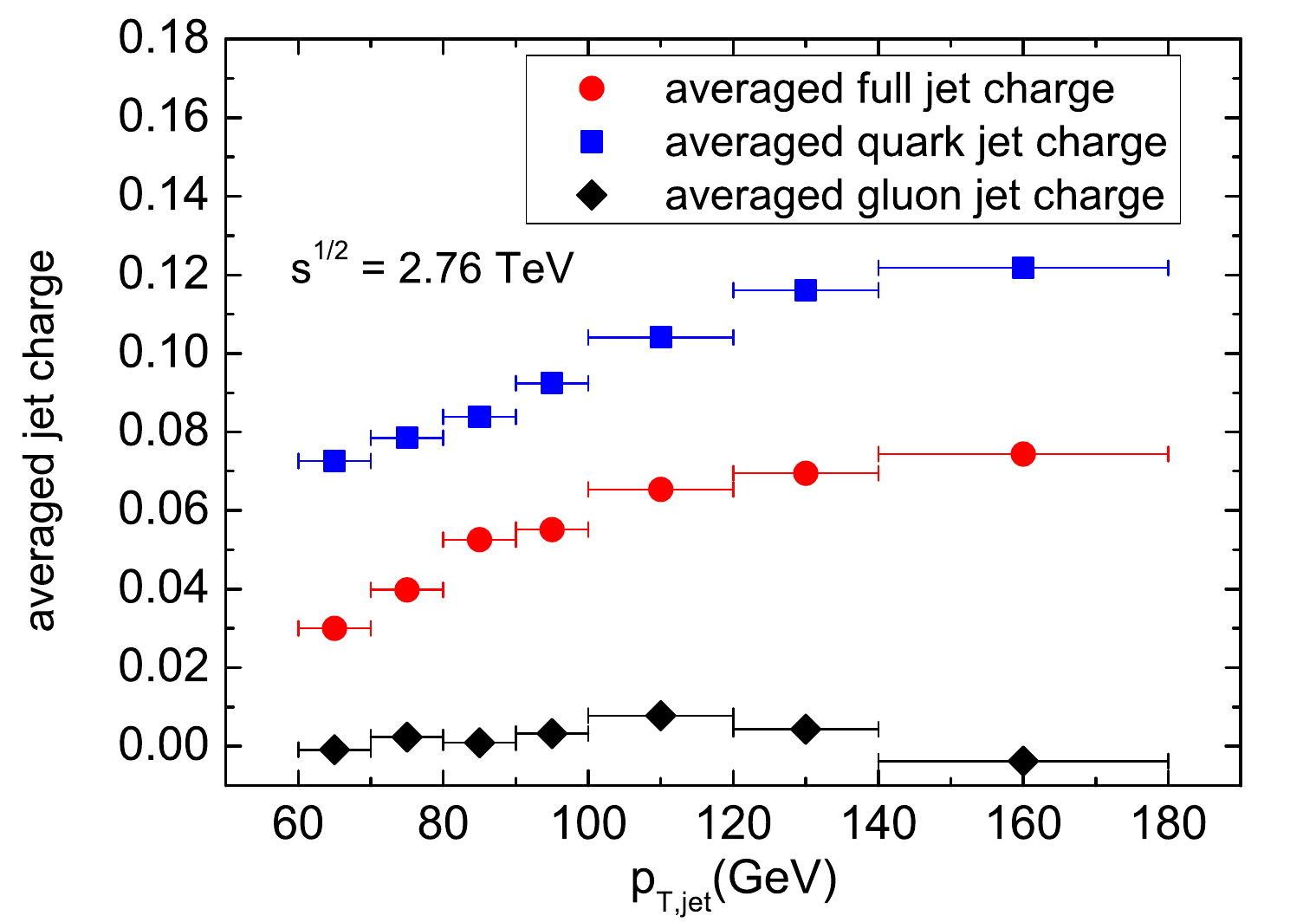}
}
\caption{The $p_T$ dependence of jet charges for quark jets, gluon jets and full jets in p+p at 2.76~TeV.}
\label{fig:qg}
\end{figure}

Moreover, we calculate the jet charge for quark and gluon jets respectively in Fig.~\ref{fig:qg}.
It is shown that the features of jet charge for quark and gluon jets are quite different. Because gluon carries no electric charge, jet charge of a gluon jet is approaching zero in a whole range of $p_T$. The averaged full jet charge
is the combination of charges of quark and gluon jets after taking into account their relative contributions.

Theoretical studies and experimental measurements
have shown that fast partons traversing the QGP medium should suffer multiple scattering with other partons and lose energy due
to collisional and radiative processes. In our calculation we will use PYQUEN(PYthia QUENched) model~\cite{Lokhtin:2005px} to simulate parton energy loss in the QGP medium.
PYQUEN is one of the Monte Carlo models
of jet quenching and built as a modification
of jet events obtained for hadronic collisions with
pythia$\_$6.4. The details of the used physics model
and simulation procedure can be found in~\cite{Lokhtin:2005px}.
The model has include both
radiative and collisional energy loss of hard partons, as well as the
realistic nuclear geometry.
The collisional energy loss is considered by the formula:

\begin{equation}
\label{collisional}
\frac{dE^{col}}{dl}=\frac{1}{4T\lambda\sigma}\int^{t_{max}}_{\mu^{2}_{D}}dt\frac{d\sigma}{dt}t.
\end{equation}
The radiative energy loss is simulated in the
frameworks of BDMPS model~\cite{Baier:1996sk,Baier:1996kr},
which gives~\cite{Lokhtin:2005px}:

\begin{equation}
\frac{dE^{rad}}{dl}=\frac{2\alpha_{s}(\mu^{2}_{D})}{\pi L}\int^{E}_{\omega_{min}}d\omega [1-y+\frac{y^{2}}{2}]
ln |\cos(\omega_{1}t_{1})| \,\, .
\end{equation}

\begin{figure}[!htb]
\centerline{
\includegraphics[width = 70mm, height=2.5in,angle=0]{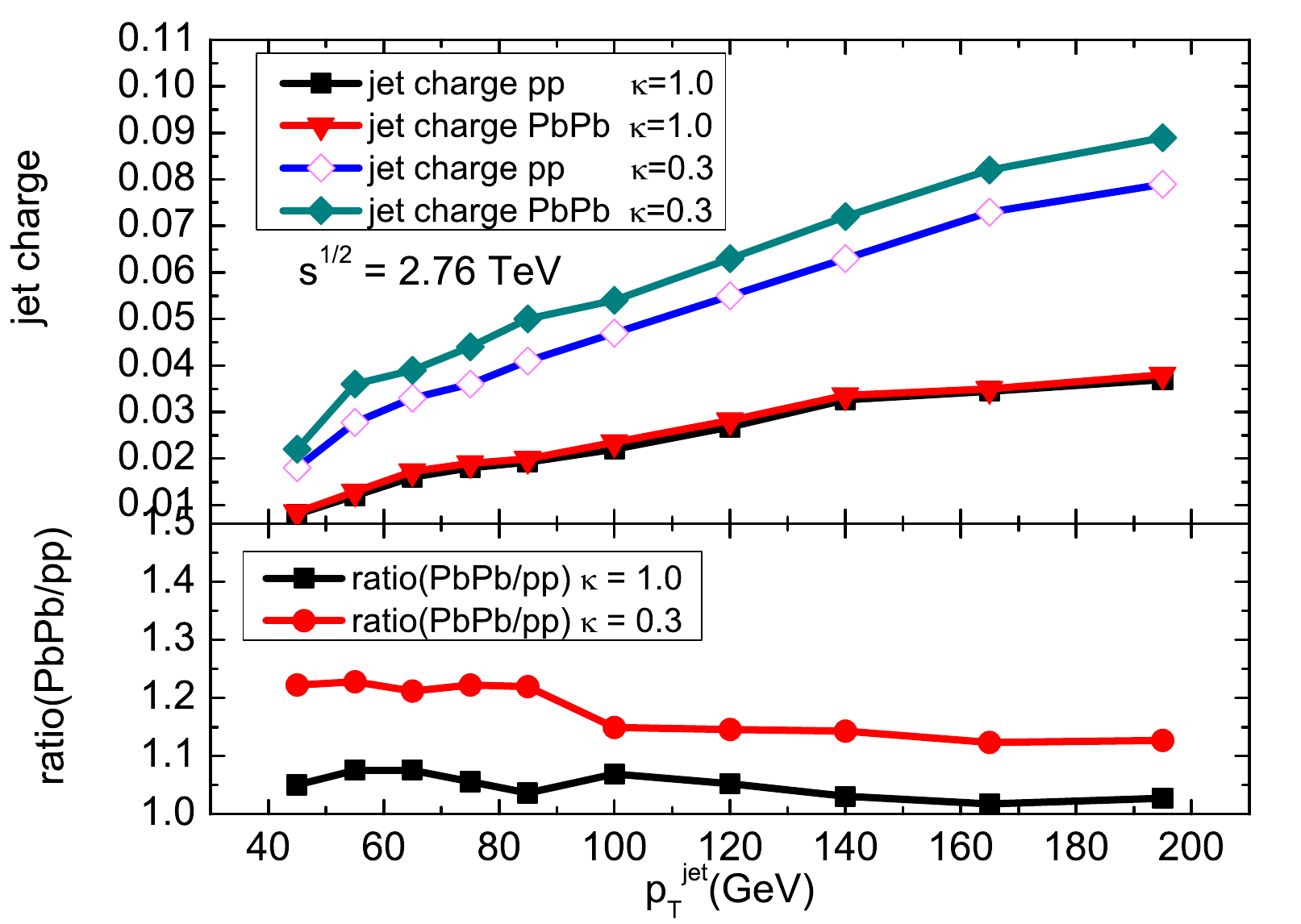}
}
\caption{The average jet charge distribution for the leading jet
in dijet events as a function of the jet transverse momentum in Pb+Pb and p+p collisions at 2.76~TeV.}
\label{fig:2.76}
\end{figure}
To investigate jet observables, another important ingredient of the model is the angular spectrum of medium-induced
radiation. It has been found that to reproduce experimental data on
jet observables such as inclusive jet cross sections, jet shape and jet fragmentation function in heavy-ion collisions, the "wide-angle" radiation scenario should be used in PYQUEN model~\cite{Lokhtin:2014vda}, which is also adopted in the calculations of this work.

It is noted that in the original version of PYQUEN model, the evolution of the QGP is described by a longitudinal expansion 1+1D Bjorken model with initial condition $\tau_{0} = 0.6$~fm and $T_{0} = 1$~GeV at central Pb+Pb collisions. In order to have a more realistic evolution of the fireball produced in relativstic heavy-ion reactions, in a modified version of PYQUEN we have instead used an event-by-event 3+1D ideal hydrodynamical model with an initial condition $\tau_{0} = 0.4$~fm and $T_{0} = 0.39$~GeV,  which has given a nice descrption on the bulk properties in Pb+Pb collsions at the LHC~\cite{Pang:2012he}.

\section{Numerical results and discussions}
\label{sec:results}
With the framework discussed in the last section, we calculate the average jet charge distribution for the leading jet
in dijet events as function of the jet transverse momentum at the LHC energy. The candidate particles used to reconstruct jets
are required to have $p_{T}>500$~MeV.
Jets are clustered by the anti-$k_{t}$ jet algorithm~\cite{Cacciari:2008gp} with radius  $R=0.4$ implemented in
FastJet. We also make constraints of jet rapidity with $|\eta|<2.1$, the two leading jets with $p^{lead}_{T}/p^{sublead}_{T}<1.5$. Here $p^{lead}_{T}$ and $p^{sublead}_{T}$ are the transverse energy of the jets with the highest and second-highest $p_{T}$, respectively.

Shown in Fig.~\ref{fig:2.76} are our theoretical predictions for jet charges with $\kappa=0.3,\,1$ in Pb+Pb collisions with $\sqrt{s_{NN}}=2.76$~TeV at the LHC, the corresponding p+p baselines, as well as their ratios~\cite{Chen:2016}. It could be seen that the jet quenching effect will increase the averaged jet charge in heavy-ion collisions and this enhancement is more pronounced with smaller $\kappa$.

This medium enhancement of jet charge in heavy-ion collisions results mainly from the suppressed gluon fraction due to jet quenching. From Fig.~\ref{fig:qg} we know that gluon jet contribution to the charge distribution of an indiscriminative full jet is negligible. In heavy-ion collisions, the gluon jet will lose more energy as compared to quark jet and its relative fraction in a full jet will be reduced in nuclear collisions.   Therefore, in heavy-ion collisions the possibility of obtaining a quark-originated jet is enhanced with respect to p+p, which leads to larger jet charge because jet charge of quark is always much larger than gluon jet.  Fig.~\ref{fig:qg} illustrates
the nuclear modifications of jet charges for both quark jet and an indiscriminative jet, and one observes that the medium modification of quark jet charge is rather small.

\begin{figure}[!htb]
\includegraphics[width =70mm, height=2.3in,angle=0]{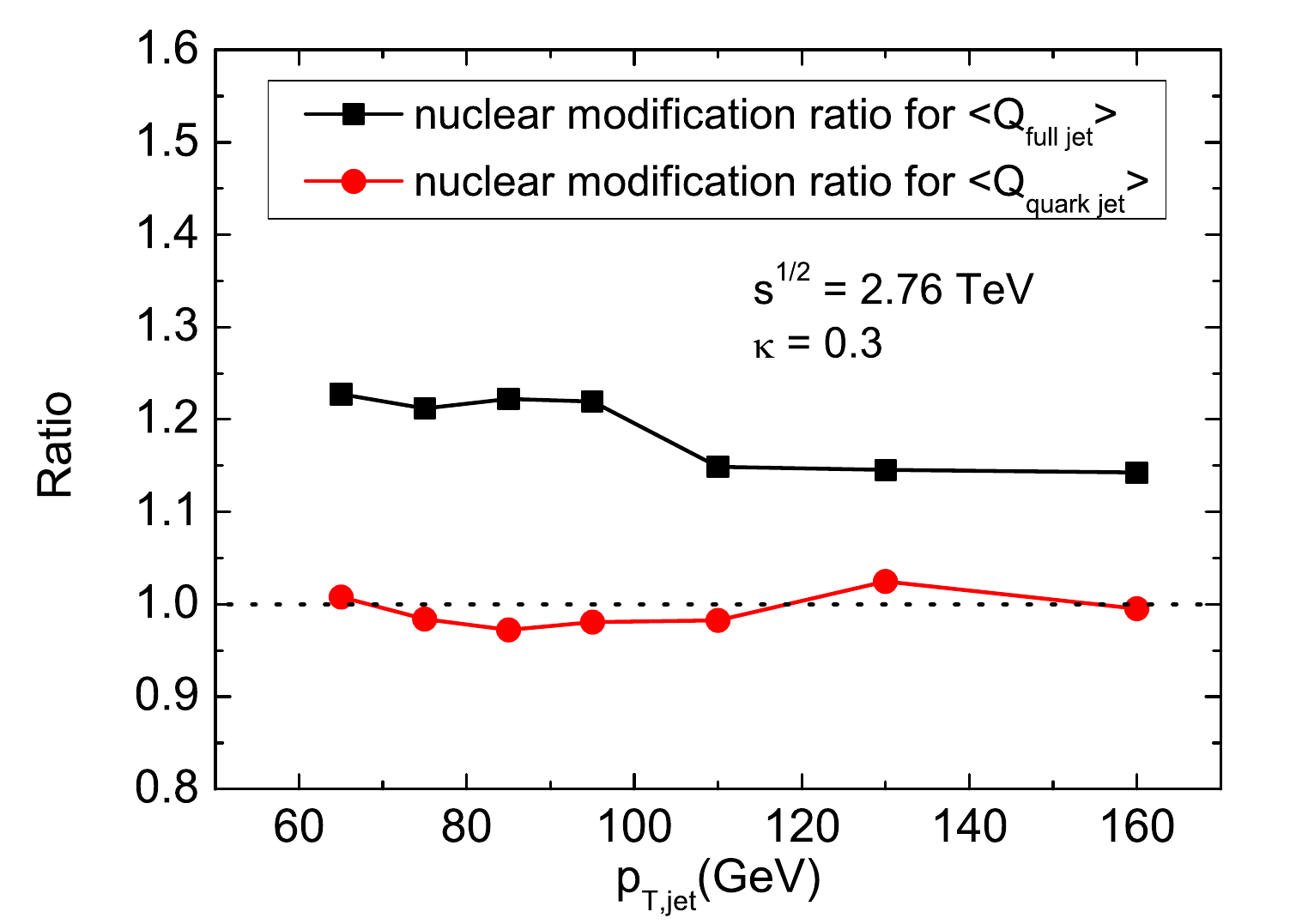}
\caption{The nuclear modification ratios of jet charges for a quark-initiated jet and an indiscriminative jet.}
\label{fig:R-q-jet}
\end{figure}

In conventional theoretical models of jet quenching, gluon is
excepted to lose more energy than quark because of its large color factor with $f=\Delta E_{g}/\Delta E_{q}=9/4$. However, the puzzle of proton over pion ratios in Au+Au collisions at RHIC seems challenge this relation~\cite{Liu:2006sf,Chen:2008vha} and our understanding of flavour dependence of parton energy loss( gluons, light quarks and massive quarks) is far from satisfactory. To investigate the sensitivity of jet charge to the different energy loss patterns of quarks and gluons we consider two simple scenarios of energy loss patterns (Scenario I with $f=9/4$ and Scenario II with $f=1$) and then calculate the corresponding jet charges.

\begin{figure}[!htb]
\includegraphics[width = 70mm, height=2.3in,angle=0]{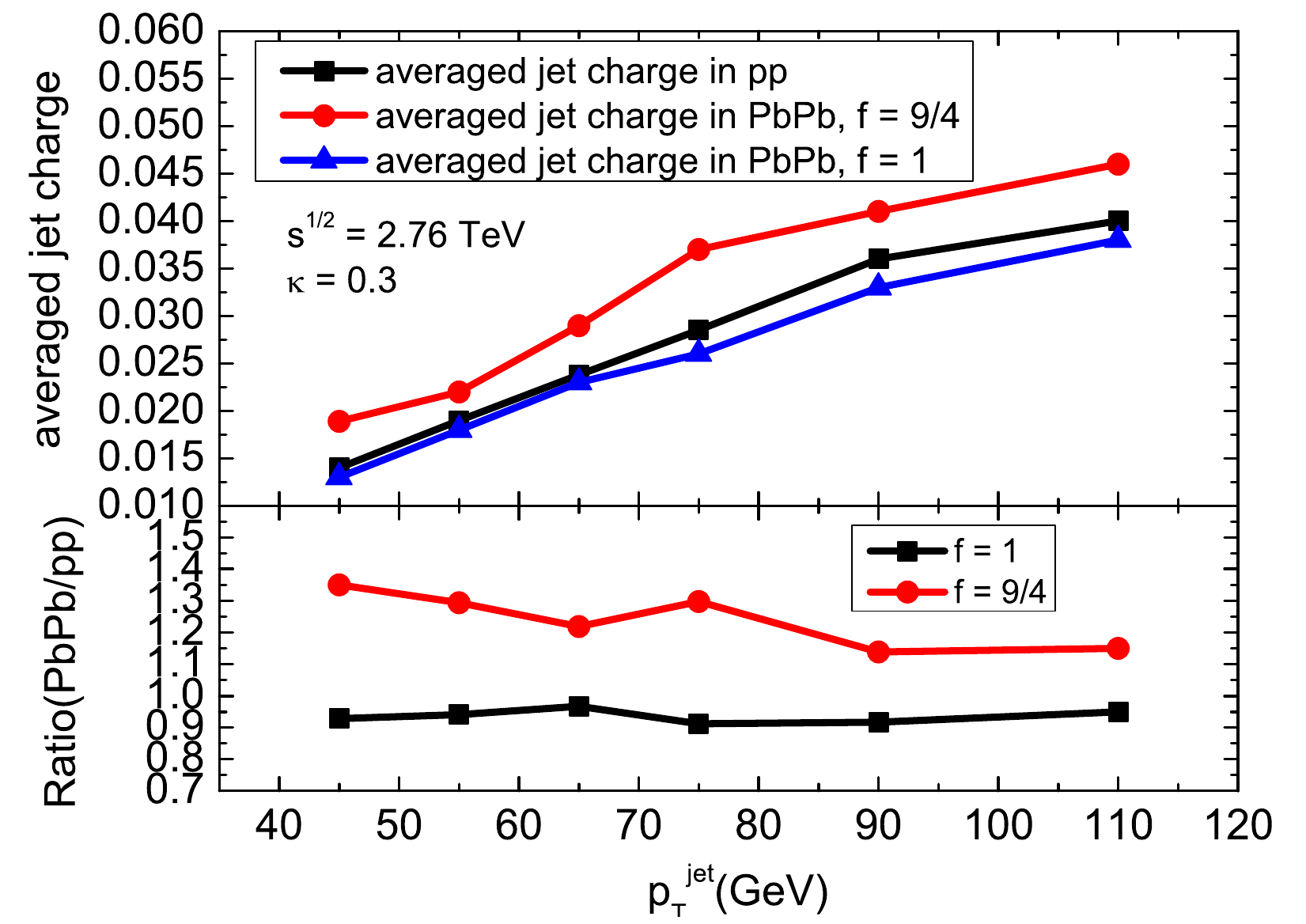}
\caption{Jet charge at Pb+Pb collisions in two scenarios of flavour dependence of parton energy loss.}
\label{fig:f1}
\end{figure}

\begin{figure}[!htb]
\includegraphics[width = 70mm,height=2.2in,angle=0]{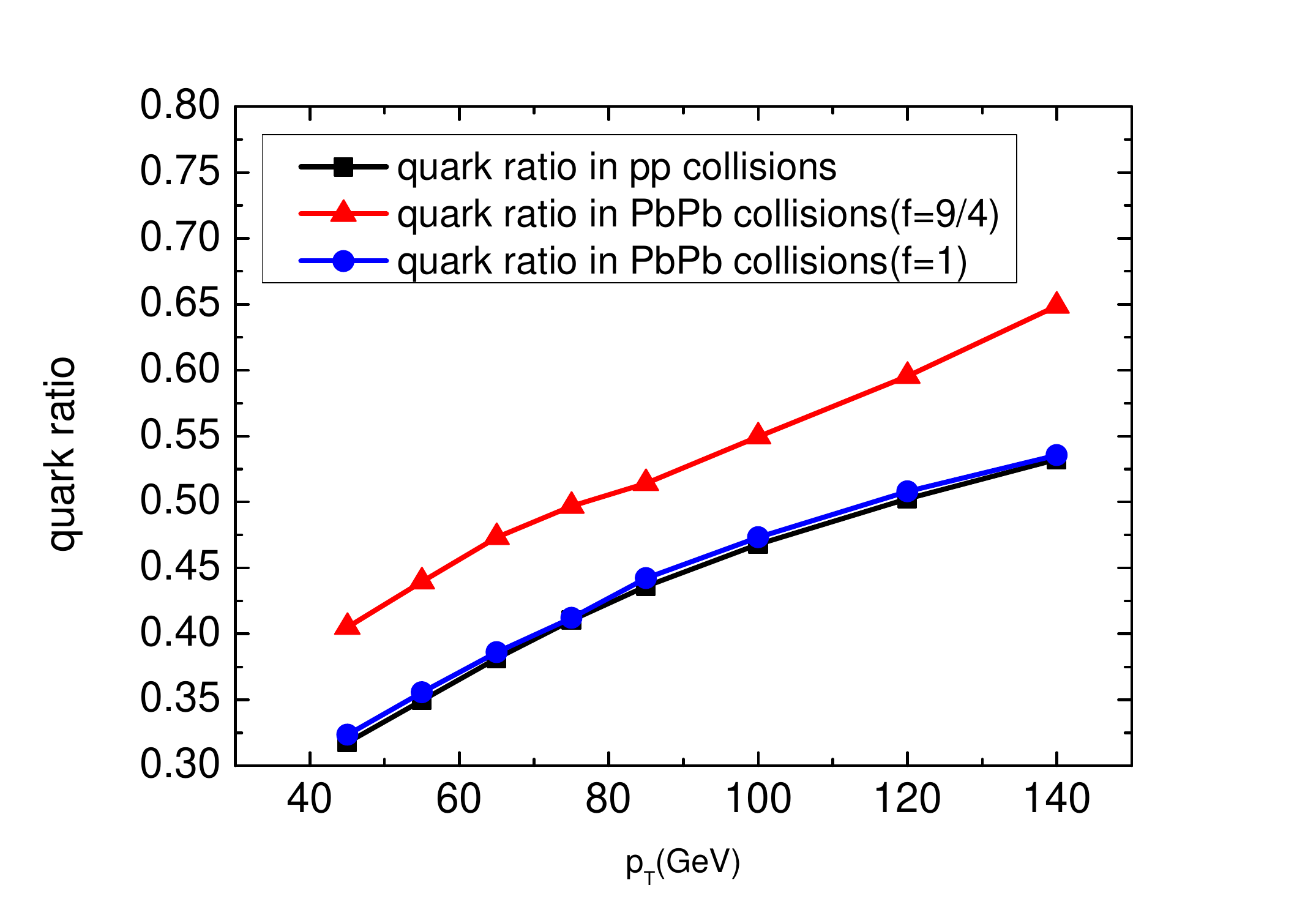}
\caption{The fraction of quark-originated jets in Pb+Pb collisions in two scenarios of flavour dependence of parton energy loss. }
\label{fig:quark-ratio}
\end{figure}

We show the numerical results in Fig.~\ref{fig:f1} and , and find that Scenario I with $f=9/4$ will give larger averaged jet charge than that in Scenario II
with $f=1$. This difference results from the enhancement of relative fraction of quark jets in Scenario I as shown in Fig.~\ref{fig:quark-ratio}, and the larger fraction of quark jets will increase jet charge because the charge of gluon-originated jet is negligible. So a comparison of the sophisticated theory and precise measurement of jet charge in heavy-ion collisions should help unravel the flavour patterns of parton energy loss.

\section{Summary}
\label{sec:summary}

We calculate numerically the momentum-weighted sum of jet charges in relativistic heavy-ion collisions for the first time.
It is shown that the jet charge is significant modified by the jet quenching effect in
the QGP medium. Quark jet charge has shown a quite distinct feature with gluon jet charge, and the sensitivity of jet charge to flavour dependence of parton energy loss is also investigated. It could be seen that jet change is an excellent observable to discriminate quark and gluon jets and their energy loss patterns in heavy-ion collisions.

We thank Prof. X.~N.~Wang, Wei Chen and G.~Y.~Ma for helpful discussions.
This research is supported by the MOST in China under Project No. 2014CB845404, and by NSFC of China with Project Nos. 11322546, 11435004, and 11521064.











\end{document}